\documentclass[journal=jctcce,manuscript=article]{achemso}

\usepackage[version=3]{mhchem} 
\usepackage{graphicx}
\usepackage{xcolor}
\usepackage{bm}
\usepackage{hyperref}
\usepackage{multirow}

\author{Kenji Sugisaki}
\affiliation[Keio University]{Graduate School of Science and Technology, Keio University, 7-1 Shinkawasaki, Saiwai-ku, Kawasaki, Kanagawa 212-0032, Japan}
\alsoaffiliation[KQCC]{Quantum Computing Center, Keio University, 3-14-1 Hiyoshi, Kohoku-ku Yokohama, Kanagawa 223-8522, Japan}
\alsoaffiliation[TCG Crest]{Centre for Quantum Engineering, Research and Education TCG Centres for Research and Education in Science and Technology, Sector V, Salt Lake, Kolkata 700091, India}
\email{ksugisaki@keio.jp}

\title[]
  {Projective Quantum Phase Difference Estimation Algorithm for the Direct Computation of Eigenenergy Gaps on a Quantum Computer}


\begin{document}

%
%
%
%
%

\begin{abstract}
 Quantum computers are capable of calculating the energy gap of two electronic states by using the quantum phase difference estimation (QPDE) algorithm. The Bayesian inference based implementations for the QPDE have been reported so far, but this approach is not projective, and the quality of the calculated energy gap depends on the input wave functions being used. Here, we report the inverse quantum Fourier transformation based QPDE with $N_a$ of ancillary qubits, which allows us to compute the difference of eigenenergies based on the single-shot projective measurement. As a proof-of-concept demonstrations, we report numerical experiments for the singlet--triplet energy gap of hydrogen molecule and the vertical excitation energies of halogen-substituted methylenes (CHF, CHCl, CF$_2$, CFCl and CCl$_2$) and formaldehyde (HCHO).   
\end{abstract}

\section{Introduction}
Quantum computing is one of the most innovative research fields in current science, and it is anticipated to bring breakthroughs in quantum chemical calculations. Quantum chemical calculations are based on the Schr{\"o}dinger equation, which governs the dynamics of quantum particles, and accurate quantum chemical calculations can potentially open the door to predictive quantum chemistry. However, as Dirac noted, ``The underlying physical laws necessary for the mathematical theory of a large part of physics and the whole of chemistry are thus completely known, and the difficulty is only that the exact application of these laws leads to equations much too complicated to be soluble.''.\cite{Dirac1929} The computational cost of the full configuration interaction (full-CI) method, which is the variationally best possible wave function in the Hilbert space spanned by the basis set used, scales exponentially with system size, and it is impractical except for small molecules with medium-size basis set. 

Since quantum chemical calculation deals with the dynamics of electrons in atoms and molecules, it is a potentially amenable problem for quantum computers. In fact, in 2005 Aspuru-Guzik and coworkers reported a method for performing the full-CI calculation on a quantum computer,\cite{AAG2005} using a quantum phase estimation (QPE) algorithm.\cite{QPE_Lloyd} The QPE is a quantum algorithm that is capable of computing the eigenvalues and corresponding eigenvectors of a unitary operator $U$. By using the time evolution operator $\exp(-iHt)$ for $U$, one can calculate the full-CI energies on a quantum computer. However, because QPE utilizes the projective measurement of the quantum state to obtain the eigenfunction and the corresponding eigenvalue, it is probabilistic, and which electronic state is obtained from the QPE depends on the overlap between the approximate wave function used as the input and the exact eigenfunction. Importantly, the QPE itself does not guarantee an exponential speedup of quantum chemical calculations,\cite{GKLChan2023} and connecting theoretical methods to generate sophisticated approximated wave function is necessary.

Quantum chemical calculations can afford to compute total energies of atoms and molecules, but total energies are generally not available from experiments. Chemical phenomena relevant to quantum chemical calculations are usually discussed in terms of the energy gaps between different geometries or electronic states. Thus, accurate prediction of the energy gaps is crucial for the practical use of quantum chemical calculations in chemistry research and development. 

Because quantum computers can use quantum superposition states as their computational resources, it is possible to compute the energy gaps directly on a quantum computer. In fact, several quantum algorithms for the direct estimation of energy gaps have been reported, such as the quantum annealing-based approach,\cite{EnergyGapQA} the quantum--classical hybrid algorithm,\cite{EnergyGapVQE} using the robust phase estimation technique,\cite{EnergyGapRPE} and the algorithms for fault-tolerant quantum computers including the Bayesian exchange coupling parameter calculator with broken-symmetry wave functions (BxB) algorithm\cite{BxB} and the Bayesian phase difference estimation (BPDE) algorithm.\cite{BPDE1st, BPDE2nd, BPDEgrad, BPDE-FSS} In addition to these methods, multiple eigenvalue estimation techniques for the simultaneous determination of the ground and the excited state energies have been reported.\cite{MultipleEigenvaluesQPE, BayesianMultiphase, MultiphaseEstimation} 
In particular, the BPDE algorithm reported by us is an extension of the Bayesian phase estimation (BPE)\cite{BPE_theory, BPE_Experiment, BPE_Quantinuum} to the quantum phase \textit{difference} estimation using the quantum superposition of two electronic states. BPDE is in principle applicable to arbitrary electronic states and it is free from the controlled-time evolution requited in conventional QPE algorithms for total energy calculations. The BPDE algorithm has been applied to the direct calculation of ionization energies, singlet--triplet energy gaps and valence excitation energies,\cite{BPDE1st} total energies by using the quantum superposition of the desired electronic state and the vacuum state,\cite{BPDE2nd} finite difference based numerical energy gradients,\cite{BPDEgrad} and relativistic energy gaps (fine-structure splitting).\cite{BPDE-FSS} The BPDE algorithm is a powerful tool for studying the energy gap of atoms and molecules, but it has two major weaknesses. (1) It requires many circuit executions to compute the energy gaps and the calculated energy gaps suffer from shot noise, and (2) it is not projective, and the calculated energy gap depends on the approximated wave functions used as the input. The second drawback can be clearly seen in the total energy calculations of the transition state between \textit{trans}- and \textit{iso}-isomers of the $\mathrm{N_2H_2}$ molecule.\cite{BPDE2nd} It is desirable to extend the quantum phase difference estimation (QPDE) algorithm to the projective methods. 

In this work, we propose a projective QPDE algorithm with $N_a$ of ancillary qubits. Hereafter we denote the proposed approach as the ``$N$-qubit QPDE'' algorithm. The $N$-qubit QPDE is a natural extension of the $N$-qubit QPE to the phase difference estimation, and it is able to compute the eigenenergy difference even if we use approximated wave functions as inputs. Similar to the $N$-qubit QPE for total energy calculations, which eigenenergy gap is obtained in the $N$-qubit QPDE depends on the overlap between the approximated wave functions and the eigenfunctions. To our knowledge, all other approaches for the direct calculation of the energy gap reported so far are neither projective nor single-shot protocols, and this is the first attempt to perform the eigenenergy gap calculation on a quantum computer without repetitive quantum circuit executions. As the demonstrations, we report numerical quantum circuit simulations for the singlet--triplet energy gap of $\mathrm{H_2}$ molecule, and the vertical excitation energies of halogen-substituted methylenes (CHF, CHCl, $\mathrm{CF_2}$, CFCl, and $\mathrm{CCl_2}$) and formaldehyde (HCHO). We also report the application of an algorithmic error mitigation (AEM)\cite{AEM} to remove the Trotter--Suzuki decomposition error and to improve the calculated energy differences. 

\section{Theory}
\begin{figure}[t]
\includegraphics{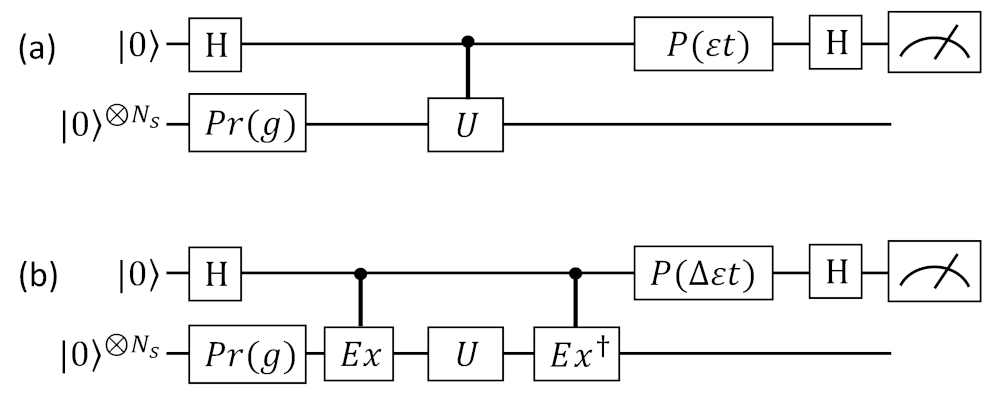}%
\caption{\label{fig:fig1} Quantum circuits for (a) the BPE and (b) the BPDE algorithms. The definition of quantum gates is given in the main text.}
\end{figure}
First, we briefly discuss the BPE\cite{BPE_theory, BPE_Experiment, BPE_Quantinuum} and the BPDE\cite{BPDE1st} algorithms. The quantum circuits for the BPE and the BPDE algorithms are shown in Figure~\ref{fig:fig1}. Here, H is an Hadamard gate, $Pr(g)$ is an approximate ground state preparation gate defined in eq~\ref{eq:one}, $U$ is a time evolution operator given in eq~\ref{eq:two}, $P$ is a phase shift gate defined in eq~\ref{eq:three}, $Ex$ is a quantum circuit that generates an approximated excited state wave function $|\Phi_1\rangle$ from the approximated ground state wave function $|\Phi_0\rangle$ as in eq~\ref{eq:four}, $\varepsilon$ and $\Delta \varepsilon$ are estimators of the total energy and the energy gap, respectively. Throughout this paper we have used $\{|\Phi\rangle\}$ for approximated wave functions and $\{|\Psi\rangle\}$ for eigenfunctions.
$N_s$ in eq~\ref{eq:one} is the number of qubits used to store the wave functions. In conventional fermion--qubit mapping methods such as the Jordan--Wigner transformation (JWT)\cite{JWT} and the Bravyi--Kitaev transformation (BKT),\cite{BKT} $N_s$ is equal to the number of spin orbitals included in the active space. Roughly speaking, the QPDE can be implemented by replacing the controlled-time evolution operation in the QPE by controlled-excitation and subsequent time evolution and controlled-deexcitation operations. 
\begin{eqnarray}
Pr(g)|0\rangle^{\otimes N_s} = |\Phi_0\rangle = \sum_j c_j |\Psi_j\rangle
\label{eq:one}
\end{eqnarray}
\begin{eqnarray}
U = \exp(-iHt)
\label{eq:two}
\end{eqnarray}
\begin{eqnarray}
P(\theta)=\left(
\begin{array}{cc}
1 & 0\\
0 & e^{i\theta}
\end{array}\right)
\label{eq:three}
\end{eqnarray}
\begin{eqnarray}
Ex|\Phi_0\rangle = |\Phi_1\rangle = \sum_k d_k |\Psi_k\rangle
\label{eq:four}
\end{eqnarray}
By expanding the approximated wave functions by eigenfunctions as given in the right hand side of eqs~\ref{eq:one} and~\ref{eq:four}, the probability of obtaining the $|0\rangle$ state in the measurement of an ancillary qubit, $Prob(0)$, can be calculated as in eqs~\ref{eq:five} and~\ref{eq:six} for the BPE and the BPDE, respectively.
\begin{eqnarray}
Prob(0; \mathrm{BPE}) = \frac{1}{2}\left[1 + \sum_j |c_j|^2 \cos\{(E_j - \varepsilon)t\}\right]
\label{eq:five}
\end{eqnarray}
\begin{eqnarray}
Prob(0; \mathrm{BPDE}) = \frac{1}{2}\left[1 + \sum_{j,k} |c_j|^2|d_k|^2 \cos\{(\Delta E_{jk} - \Delta \varepsilon)t\}\right]
\label{eq:six}
\end{eqnarray}
These equations insist that if the approximated wave functions $|\Phi\rangle$ have sufficiently large overlaps with the eigenfunctions of the target electronic states, $Prob(0)$ becomes maximum at $\varepsilon = E_j$ and $\Delta \varepsilon = \Delta E_{jk}$ for the BPE and the BPDE, respectively. In the BPE and the BPDE algorithms, total energies and energy differences are computed by searching for the $\varepsilon$ and $\Delta \varepsilon$, respectively, that give maximum $Prob(0)$, using Bayesian inference. 

From eqs~\ref{eq:five} and~\ref{eq:six}, it is clear that the accuracy of the calculated total energies and energy gaps in the framework of the BPE and the BPDE algorithms strongly depends on the quality of the approximated wave functions being used as the inputs. If the approximated wave functions are expressed by linear combinations of many eigenstates, and contributions from electronic states other than the target state to $Prob(0)$ become non-negligible, the peak position of $Prob(0)$ may be shifted from that corresponding to the true eigenvalues. For typical closed-shell singlet molecules in their equilibrium geometries, the Hartree--Fock wave function $|\Phi_{\mathrm{HF}}\rangle$ is a reasonably good approximation of the ground state wave function. However, the overlap squared value $|\langle\Phi_{\mathrm{HF}}|\Psi_{\mathrm{full-CI}}\rangle|^2$ can decrease exponentially with the system size, and therefore using $|\Phi_{\mathrm{HF}}\rangle$ as the input wave function is only valid for small molecules. In addition, the excited state wave function is usually more complicated than that of the ground state, hence generating a good approximated wave function for the excited state is in general a difficult task. One possible solution is to adopt an adiabatic state preparation (ASP) algorithm to generate correlated wave functions.\cite{AAG2005, ASP_Veis, ASP_Kremenetski, ASP_CommsChem} However, the quantum circuit for the ASP is usually deep and it is unsuitable for the BPE and BPDE frameworks which need repetitive quantum circuit executions. 

\begin{figure}[tbp]
\includegraphics{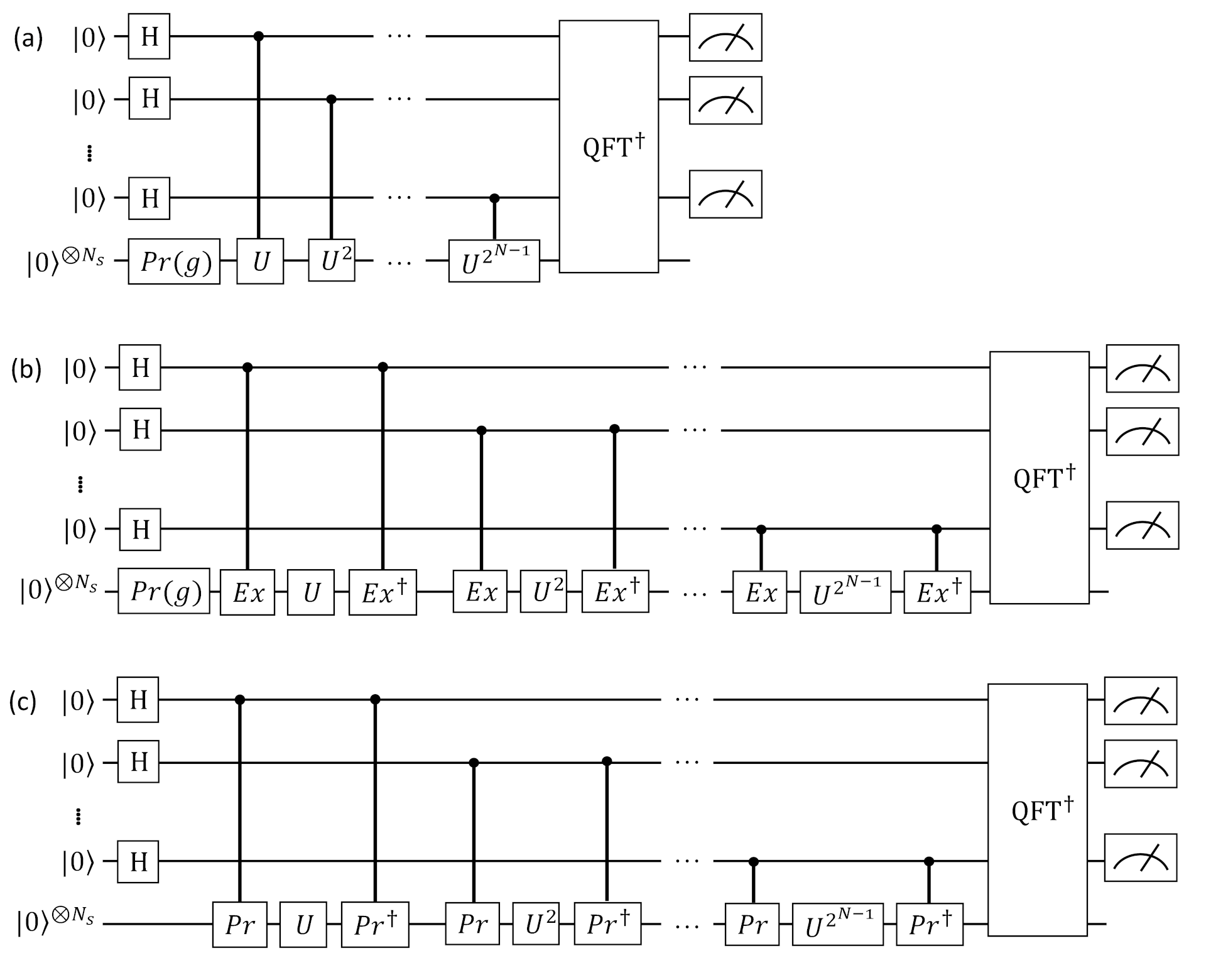}%
\caption{\label{fig:fig2} Quantum circuits for (a) the $N$-qubit QPE algorithm, (b) the $N$-qubit QPDE algorithm in a naive implementation, and (c) the $N$-qubit QPDE algorithm used in this work.}
\end{figure}
In the total energy calculations, single-shot projective calculations are possible by using the QPE algorithm with $N_a$ of ancillary qubits.\cite{AAG2005} The quantum circuit for the $N$-qubit QPE is shown in Figure~\ref{fig:fig2}(a). Here, QFT$^{\dagger}$ represents the quantum circuit for the inverse quantum Fourier transformation.\cite{NielsenChuang} By assuming eq~\ref{eq:one}, the quantum state before the measurement of ancillary qubits in Figure~\ref{fig:fig2}(a) is calculated as in eq~\ref{eq:seven}. 
\begin{eqnarray}
|0\rangle^{\otimes N_a} \otimes |0\rangle^{\otimes N_s} \xrightarrow{\mathrm{QPE}} \sum_j c_j |\phi_j\rangle \otimes |\Psi_j\rangle 
\label{eq:seven}
\end{eqnarray}

Thus, the measurement of ancillary qubits yields an eigenphase $\phi_j = 0.x_1x_2x_3...x_{N_s}$ in binary fraction with the probability proportional to $|c_j|^2$, and the wave function is projected onto the corresponding eigenfunction $|\Psi_j\rangle$. Here, $\{x_j\}$ are the measurement outcome of the $j$-th ancillary qubit. The eigenenergy $E$ can be calculated by using the equation $E = -2\pi\phi/t$, which is derived from the equation $e^{-iHt}|\Psi\rangle = e^{-iEt}|\Psi\rangle = e^{2i\pi\phi}|\Psi\rangle$. From an analogy of the extension of the BPE algorithm to the BPDE, we can construct the quantum circuit for the $N$-qubit QPDE algorithm by replacing the controlled-$U$ gates with the sequence of controlled-$Ex$, $U$, and controlled-$Ex^\dagger$ gates, as in Figure~\ref{fig:fig2}(b). Note that the quantum circuit in Figure~\ref{fig:fig2}(b) contains control-free time evolution operations, and therefore this implementation implies that $|\Phi_0\rangle$ is an eigenfunction of the time evolution operator. Unfortunately, this assumption is not generally true. This difficulty can be avoided by starting from the $|0\rangle^{\otimes N_s}$ state and using the controlled-$Pr$ gate given in eq~\ref{eq:eight} instead of using the controlled-$Ex$ gate, as shown in Figure~\ref{fig:fig2}(c).  
\begin{eqnarray}
\textrm{controlled-}Pr = |0\rangle \langle0|\otimes Pr(g) + |1\rangle \langle 1|\otimes Pr(e)
\label{eq:eight}
\end{eqnarray}
Here, $Pr(g)|0\rangle^{\otimes N_s} = |\Phi_0\rangle$ and $Pr(e)|0\rangle^{\otimes N_s} = |\Phi_1\rangle$. 
Since the $|0\rangle^{\otimes N_s}$ state in JWT and BKT corresponds to the vacuum state with no electrons, applying $U$ does not change the state, $U|0\rangle^{\otimes N_s} = |0\rangle^{\otimes N_s}$. As a result, the quantum circuit for the $N$-qubit QPDE is slightly deeper than that for the $N$-qubit QPE. Nevertheless, the $N$-qubit QPDE-based direct calculation of the energy gap is still advantageous than the putative approach based on the $N$-qubit QPE for total energy calculations of individual electronic states, as long as the controlled-$Pr$ gate can be realized by shallow quantum circuit. In the present study, the approximated wave functions are expressed by a linear combination of at most two Slater determinants, and thus the quantum circuits for controlled-$Pr$ are sufficiently shallow. 

In the implementation described in Figure~\ref{fig:fig2}(c), the bit string obtained from the measurement of ancillary qubits corresponds to the eigenphase difference $\Delta \phi_{jk} = 0.x_1x_2x_3...x_{N_s}$, and the eigenenery gap can be calculated as $\Delta E_{jk} = -2\pi\Delta\phi_{jk}/t$. The probability of which eigenenergy gap is obtained is proportional to $|c_j|^2\times|d_k|^2$. 

It should be noted that the energy gap can be positive or negative, but the $\Delta \phi_{jk}$ obtained from the $N$-qubit QPDE quantum circuit does not contain information about the digits above the arithmetic point. Therefore, $\Delta \phi_{jk}$ and $\Delta \phi_{jk} - k$, where $k$ is an arbitrary integer, cannot be distinguished in the $N$-qubit QPDE. This means that there is a possibility to assign a wrong sign to the energy gap. To avoid such a misassignment, we can appropriately set the evolution time length $t$ so that $\cos(2\pi\Delta \phi_{jk}) > 0$, and use $\Delta \phi_{jk}$ when $0 \le \Phi_{jk} \le 1/4$ and $\Delta \phi_{jk} - 1$ when $3/4 \le \phi_{jk} < 1$. 

\section{Computational conditions}
As a proof-of-concept demonstration of the proposed $N$-qubit QPDE algorithm, here we report numerical simulations for the direct calculation of the singlet--triplet energy gap of H$_2$ molecule, and the vertical excitation energies of halogen-substituted methylenes (CHF, CHCl, CF$_2$, CFCl, and CCl$_2$) and formaldehyde (HCHO). We computed the singlet--triplet energy gap as the excitation energy of the spin-triplet state from the spin-singlet state. Because all the molecules under study have the spin-singlet ground state, the singlet--triplet energy gap $\Delta E_{\mathrm{ST}} = E(\mathrm{T_1}) - E(\mathrm{S_0})$ is positive. 
Note that numerical simulations of the BxB and the BPDE-based energy gap calculations have been reported\cite{BxB, BPDE1st} for some of the compounds under study. In our previous publications, the singlet--triplet energy gap of H$_2$ was defined as $\Delta E = E(\mathrm{S_0}) - E(\mathrm{T_1})$, and therefore definition of the energy gap is different. 
The one- and two-electron integrals required to construct the electronic Hamiltonian were computed by using the GAMESS-US program package.\cite{GAMESS} The full-CI and the CAS-CI calculations as the references were also performed with GAMESS-US. Numerical quantum circuit simulations were performed by using our in-house Python code developed with OpenFermion\cite{OpenFermion} and Cirq\cite{Cirq} libraries.  

The singlet--triplet energy gap of H$_2$ molecule with the atom--atom distance from 0.7 to 3.0 $\mathrm{\AA}$ was studied by using the 6-31G basis set. The wave function is encoded in 8 qubits using JWT, and 12 ancillary qubits were used for the phase readout. The evolution time length in the $U = exp(-iHt)$ operator was set to $t = 10$, and the second-order Trotter--Suzuki decomposition given in eq~\ref{eq:nine} was adopted to construct the quantum circuit. We examined three different time lengths for the single Trotter steps, $\Delta t = t/M = 0.5, 1.0$, and 1.25. 
\begin{eqnarray}
e^{-iHt} = e^{-i(\sum_j \omega_j P_j)t} = \left[\prod_{j=1}^J e^{-i\omega_j P_j t/2M} \prod_{j=J}^1 e^{-i\omega_j P_j t/2M}\right]^M + O(t(\Delta t)^2)
\label{eq:nine}
\end{eqnarray}
Here, $P_j$ is a Pauli string described by a direct product of Pauli operators $\{I, X, Y, Z\}$, and $\omega_j$ is the corresponding coefficient. It is known that the Trotter--Suzuki decomposition error depends on the term ordering.\cite{TrotterError_JCTC, TrotterError_Entropy} In this study we used a magnitude ordering\cite{TrotterError_JCTC} where the Pauli string is applied in decreasing order of $|\omega_j|$. 

\begin{figure}[tbp]
\includegraphics{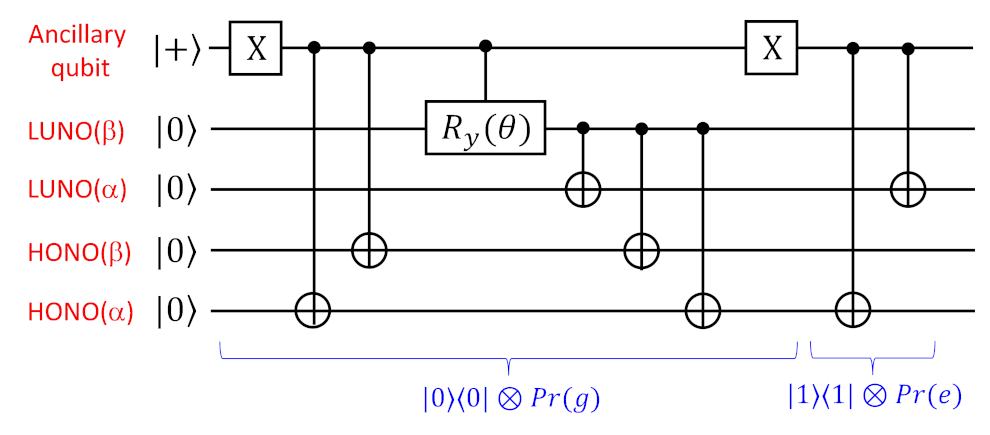}%
\caption{\label{fig:fig3} The controlled-$Pr$ gate used for the $N$-qubit QPDE calculations of H$_2$ and halogen-substituted methylenes. Qubits corresponding to doubly occupied (HONO$-$1 and below) and unoccupied (LUNO$+$1 and above) orbitals are omitted for clarity. The Pauli-X gate should be applied to the qubits corresponding to the doubly occupied orbitals those do not appear in the Figure. The rotation angle of the $R_y$ gate is set to be $\theta = -2\arccos{\sqrt{1 - y^2/2}}$, where $y$ is the diradical character calculated using eqn~\ref{eq:eleven}.}
\end{figure}

As discussed in the previous section, the probability of which energy gap can be obtained is proportional to $|c_j|^2\times|d_k|^2$, where $c_j$ and $d_k$ are the coefficients defined in eqs~\ref{eq:one} and~\ref{eq:four}, respectively. This means that the accuracy of the approximated wave functions only affects the success probability, and if success, the energy gap obtained from the $N$-qubit QPDE does not depend on the quality of the wave functions. To study the input wave function dependence in H$_2$ molecule with the 6-31G basis set, we used two different wave functions for the ground state spin-singlet wave functions. One is the Hartree--Fock wave function $|\Phi_{\mathrm{HF}}\rangle$, and the other is the two-configurational wave function $|\Phi_{2c}\rangle$ constructed by using a diradical character $y$. As reported in our precedent paper,\cite{MRwf_dircha} the $|\Phi_{2c}\rangle$ defined as in eq~\ref{eq:ten} can be a qualitatively good approximated wave function of diradical systems with a large overlap with the full-CI wave function. The diradical character $y$ was calculated from the occupation number of the lowest unoccupied natural orbital ($n_{\mathrm{LUNO}}$) constructed from the spin-unrestricted broken-symmetry Hartree--Fock wave function for the $M_S = 0$ state, as given in eq~\ref{eq:eleven}.\cite{DirChaCalc} 
\begin{eqnarray}
|\Phi_{2c}\rangle = \sqrt{1 - \frac{y}{2}}|2000\rangle - \sqrt{\frac{y}{2}}|0200\rangle
\label{eq:ten}
\end{eqnarray}
\begin{eqnarray}
y = 1 - \frac{2(1 - n_{\mathrm{LUNO}})}{1 + (1 - n_{\mathrm{LUNO}})^2}
\label{eq:eleven}
\end{eqnarray}
Here, $|2000\rangle$ is the Hartree--Fock configuration and $|0200\rangle$ is the HONO--LUNO two-electron excited determinant from the Hartree--Fock configuration in the H$_2$ molecule with 6-31G basis set (4 molecular orbitals and 2 electrons). For the spin-triplet state we used the spin-restricted open-shell Hartree--Fock configuration $|\Phi_1\rangle = |\alpha\alpha00\rangle$, where $\alpha$ indicates that the molecular orbital is singly occupied by a spin-$\alpha$ electron. The controlled-$Pr$ gates used for the calculations are depicted in Figure~\ref{fig:fig3}. 

The numerical simulations of halogen-substituted methylenes were carried out at the CAS-CI/6-31G* level, using the B3LYP/6-31G* optimized geometry. Geometry optimizations were performed using Gaussian 09 software.\cite{Gaussian09} 
We used (6e, 4o) active space for CHF and CHCl, and (10e, 6o) for CF$_2$, CFCl, and CCl$_2$, and the vertical excitation energies of the lowest spin-triplet excited state ($\mathrm{1\ ^3B_1}$ state) were calculated. Here, (\textit{k}e, \textit{l}o) specifies that the active space contains \textit{k}-electrons and \textit{l}-molecular orbitals. 
We used ten ancillary qubits for the phase readout, and the evolution time length was set to be $t = 10$. The RHF and ROHF-like single determinant wave functions were used for $|\Phi_0\rangle$ and $|\Phi_1\rangle$, respectively, and the quantum circuit given in Figure~\ref{fig:fig3} with $\theta = 0$ is used for the controlled-$Pr$ gate. 

For the direct calculations of the vertical excitation energies of formaldehyde (HCHO), we focused on the three low-lying excited states ($\mathrm{1\ ^1A_2, 1\ ^1B_1}$, and $\mathrm{2\ ^1A_1}$ states). We used the same geometry and active space as our previous publication.\cite{BPDE1st} The CAS-CI active space consists of 5a$_1$ (C--O $\sigma$), 1b$_1$ (C--O $\pi$), 2b$_2$ (in-plane 2p lone pair of O), 2b$_1$ (C--O $\pi^*$), and 9a$_1$ (C--O $\sigma^*$) orbitals. 
The excited state wave functions are approximated by the spin symmetry adapted ($\mathrm{2b_2 \rightarrow 2b_1}$), ($\mathrm{5a_1 \rightarrow 2b_1}$), and ($\mathrm{1b_1 \rightarrow 2b_1}$) one-electron excitations from the $|\Phi_{\mathrm{HF}}\rangle$ for the $\mathrm{1\ ^1A_2, 1\ ^1B_1}$, and $\mathrm{2\ ^1A_1}$ states, respectively. 
The simulations were carried out using ten ancillary qubits and $t = 10$.

\section{Results and discussion}
\textbf{Effect of input wave function in H$_2$ molecule.}  First, we checked the effect of the approximated wave functions on the success probability of the $N$-qubit QPDE and the calculated eigenphase difference value, by using $|\Phi_{\mathrm{HF}}\rangle$ and $|\Phi_{2c}\rangle$ as the input wave functions for the spin-singlet electronic ground state of H$_2$ molecule. We used two different geometries $R\mathrm{(H-H)}$ = 2.0 and 3.0 $\mathrm{\AA}$ as the representative examples of an intermediate bond-breaking and a bond-dissociation regions, respectively. The ground state wave function exhibits sizable diradical character ($y$ = 0.4398 and 0.8648 for $R\mathrm{(H-H)}$ = 2.0 and 3.0 $\mathrm{\AA}$, respectively), and therefore $|\Phi_{2c}\rangle$ is a much better approximation for the ground state wave function than $|\Phi_{\mathrm{HF}}\rangle$. 

\begin{figure}[tbp]
\includegraphics{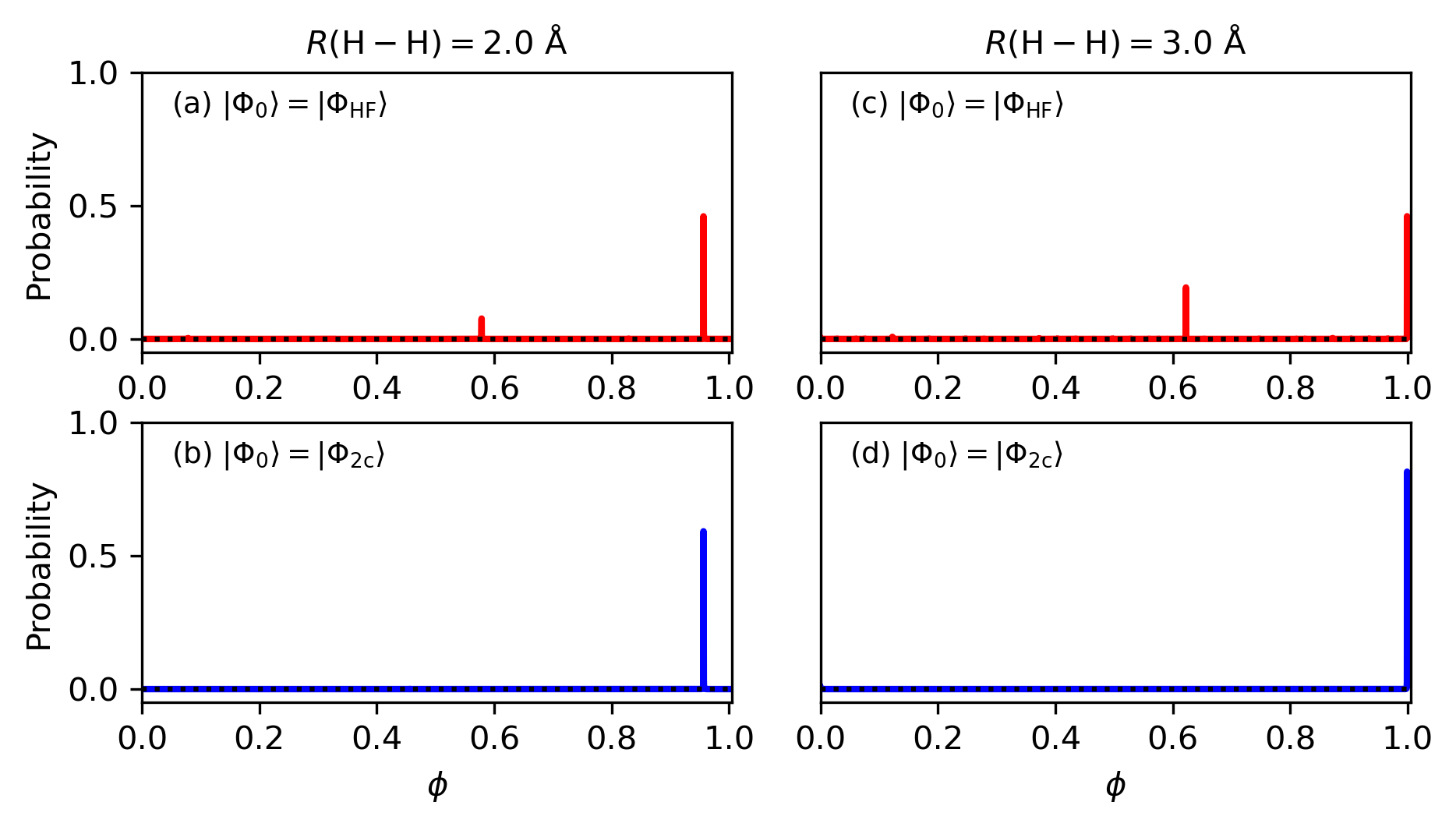}%
\caption{\label{fig:fig4} The phase value vs. probability plot obtained from the $N$-qubit QPDE simulations of H$_2$ molecule with different input wave functions for the S$_0$ ground state. }
\end{figure}

Figure~\ref{fig:fig4} summarizes the relationship between the phase value obtained from the measurement of the ancillary qubits and the probability of occurrence. The plots have two major peaks when the $|\Phi_{\mathrm{HF}}\rangle$ is used as the input wave function for the singlet ground state. By converting the phase values of the peaks at the geometry $R\mathrm{(H-H)}$ = 2.0 $\mathrm{\AA}$ to the energy unit, we obtained the energy gaps as 0.023163 and $-0.363553$ Hartree for the major and minor peaks, respectively. These values correspond to the $\mathrm{(T_1 - S_0)}$ and the $\mathrm{(T_1 - S_2)}$ energy gaps ($\Delta E$(full-CI/6-31G) = 0.027980 and $-0.361740$ Hartree, respectively). Importantly, the difference in the input wave functions only affects the success probability, and the phase value giving a peak remains unchanged. These results support the projective nature of the $N$-qubit QPDE algorithm. 

\begin{figure}[tbp]
\includegraphics{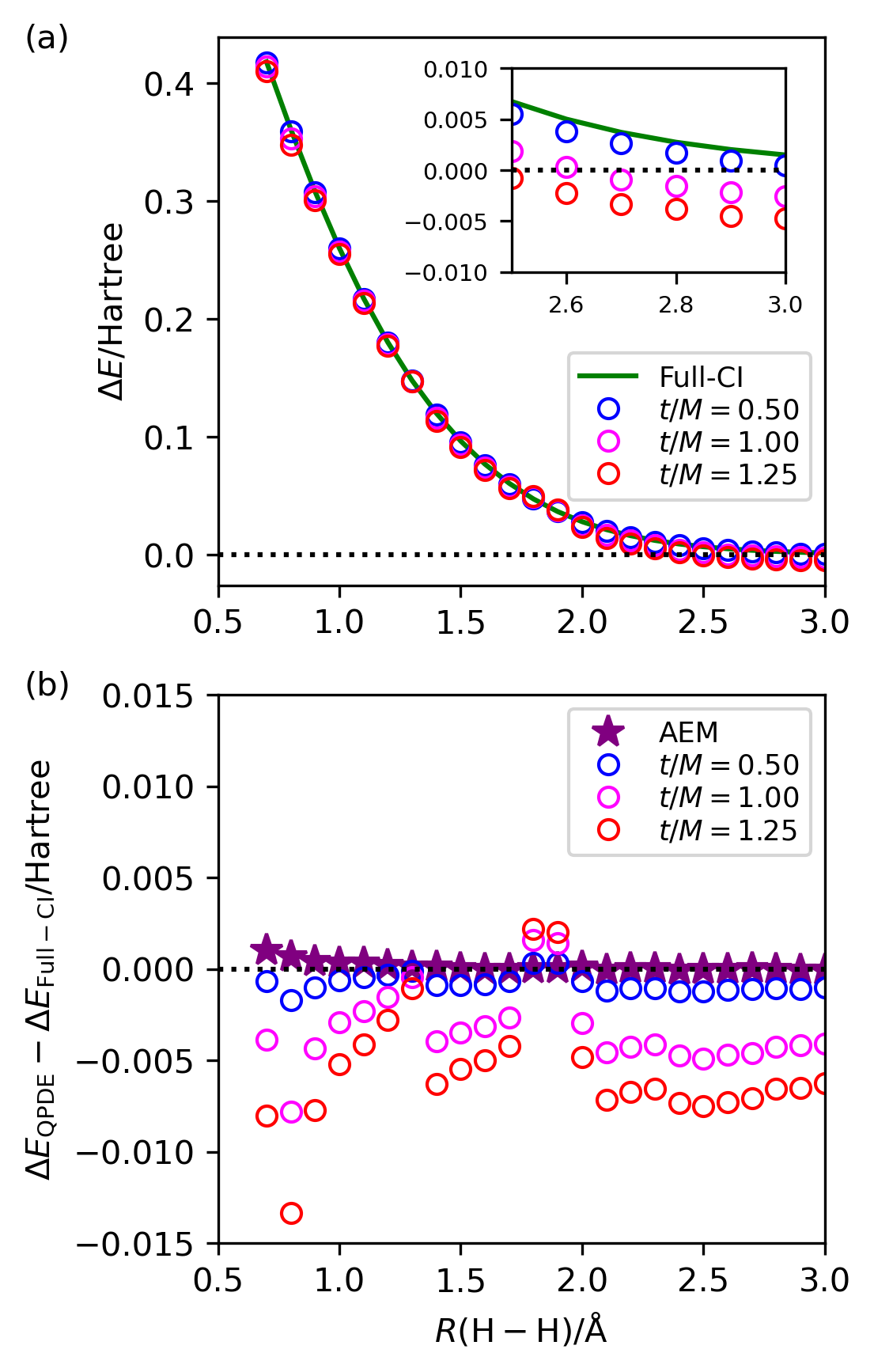}%
\caption{\label{fig:fig5} The $N$-qubit QPDE simulation results of H$_2$ molecule with different time lengths of the single Trotter step $t/M$. (a) The singlet--triplet energy gap. Inset is the plot of $\Delta E$ at the geometry $R(\mathrm{H-H}) \ge 2.5\ \mathrm{\AA}$. (b) Differences between the singlet--triplet energy gaps from the quantum circuit simulations and from the full-CI/6-31G calculations. AEM stands for the energy gap calculated using the algorithmic error mitigation technique (see text). } 
\end{figure}

\textbf{The singlet--triplet energy gap of H$_2$ molecule with different bond lengths and application of the algorithmic error mitigation.} Next, we carried out the $N$-qubit QPDE simulations for the direct calculation of the singlet--triplet energy gap of H$_2$ molecule by changing the atom--atom distance from 0.7 to 3.0 $\mathrm{\AA}$. The calculated energy gaps are plotted in Figure~\ref{fig:fig5}(a) and the deviations from the full-CI/6-31G energy gap are given in Figure~\ref{fig:fig5}(b). Here, we used the $|\Phi_{2c}\rangle$ as the input wave function of the spin-singlet state, and the phase value giving the maximum measurement probability was taken as the eigenphase. From Figure~\ref{fig:fig5}, it is clear that the deviation from the full-CI value becomes smaller when shorter time length is adopted for the single Trotter step. For the geometry with longer $R\mathrm{(H-H)}$ distances, the $N$-qubit QPDE simulations with $t/M 
= 1.00$ and 1.25 gave negative $\Delta E$ values (see inset of Figure~\ref{fig:fig5}(a)), which are in contradiction with the full-CI/6-31G results. 

The plot in Figure~\ref{fig:fig5}(b) indicates that the $\Delta E$ values obtained from the $N$-qubit QPDE simulations have systematic errors. We assume that the main source of the errors is the Trotter--Suzuki decomposition. Theoretically, using a shorter time length for the single Trotter step can systematically reduce the error, but there is a trade-off between the Trotter--Suzuki decomposition error and the computational cost. Instead of using a shorter time length for the single Trotter step, here we examined an algorithmic error mitigation (AEM) method.\cite{AEM} The AEM is a technique for mitigating the errors of algorithmic origin such as the Trotter--Suzuki decomposition. It is known that the error in the second-order Trotter--Suzuki decomposition scales as $O(t(\Delta t)^2)$, as given in eq~\ref{eq:nine}. Since we have fixed the evolution time length $t$ for different $\Delta t = t/M$ simulations, the energy gap including the Trotter--Suzuki decomposition error can be fitted by a quadratic function, $f(\Delta t) = a(\Delta t)^2 + b$, where $f(\Delta t)$ is the energy gap obtained from the $N$-qubit QPDE simulation with the single Trotter step size $t/M = \Delta t$. The residue $b$ of the fitted function corresponds to the energy gap of the Trotter--Suzuki decomposition error zero limit estimated from the extrapolation. The results of the AEM have also been plotted in Figure~\ref{fig:fig5}(b) (purple stars). It is clear that the AEM efficiently reduces the error in the energy gap estimation. In the case of H$_2$ molecule, the maximum value of $|\Delta E\mathrm{(AEM)} - \Delta E(\mathrm{full}$-CI)$|$ is 0.00101 Hartree = 0.633 kcal mol$^{-1}$, which is below the chemical precision (1.0 kcal mol$^{-1}$). 

\begin{figure}[tbp]
\includegraphics{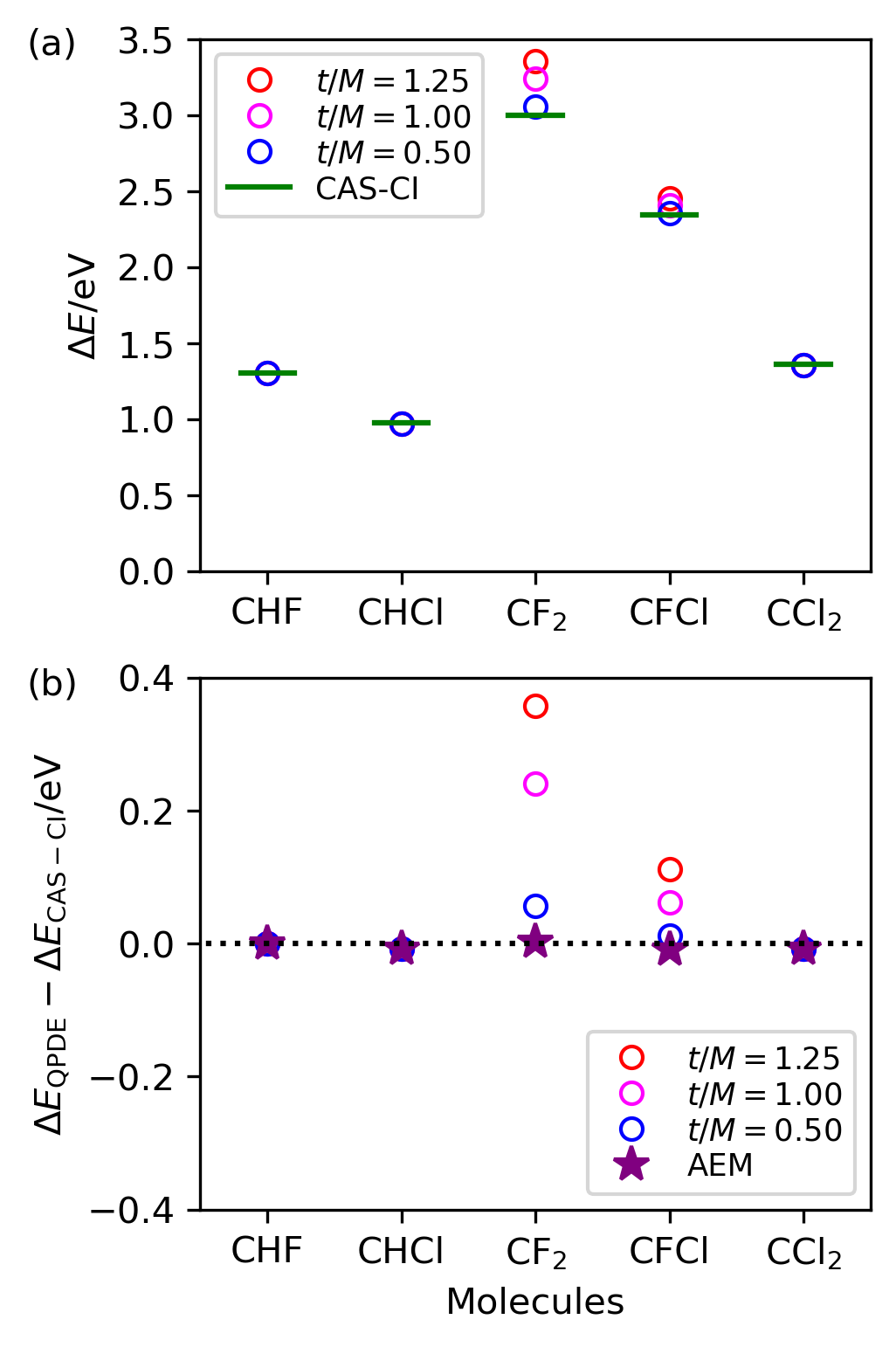}%
\caption{\label{fig:fig6} The $N$-qubit QPDE simulation results of the vertical lowest triplet excitation energies of halogen-substituted methylenes with different time lengths of the single Trotter step $t/M$. (a) The excitation energy. (b) Differences between the excitation energies from the quantum circuit simulations and those from the CAS-CI/6-31G* calculations. AEM stands for the energy gap obtained by applying the algorithmic error mitigation.} 
\end{figure}

\textbf{Vertical lowest-triplet excitation energies of halogen-substituted methylenes.} Next, we focused on the vertical excitation energy of the lowest spin-triplet state ($\mathrm{1\ ^3B_1}$ state) of halogen-substituted methylenes (CHF, CHCl, CF$_2$, CFCl, and CCl$_2$). Note that in the precedent paper we reported the direct calculation of the vertical excitation energies of CF$_2$ and CCl$_2$ molecules using the BPDE algorithm.\cite{BPDE1st} 

The results of the $N$-qubit QPDE simulations are plotted in Figure~\ref{fig:fig6}. We observed strong system dependence of the Trotter--Suzuki decomposition errors. In CHF, CHCl, and CCl$_2$ molecules, the $N$-qubit QPDE simulations with three different $t/M$ values yielded the same energy gap, and the calculated energy gaps agree with the CAS-CI values with less than 0.01 eV of deviation. In contrast, the Trotter--Suzuki decomposition errors are significant for CF$_2$ and CFCl molecules. For example, the deviations of the excitation energies obtained from the $N$-qubit QPDE from the CAS-CI values in CF$_2$ are 0.056, 0.240, and 0.357 eV for $t/M$ = 0.50, 1.00, and 1.25, respectively. However, by adopting the AEM to mitigate the Trotter--Suzuki decomposition error, the departure of the excitation energy from the CAS-CI value becomes 0.002 eV. This result also illustrates the importance of the AEM in accurately predicting excitation energies using the $N$-qubit QPDE algorithm. 

\begin{figure}[tbp]
\includegraphics{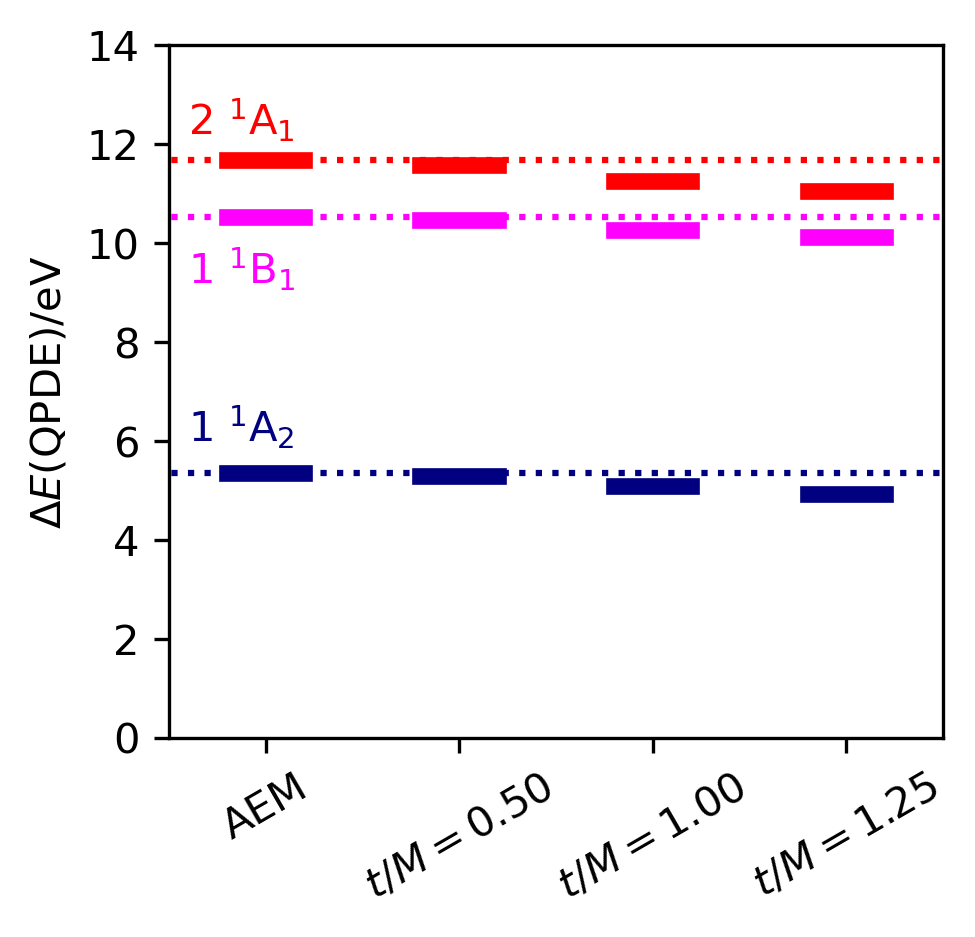}%
\caption{\label{fig:fig7} The $N$-qubit QPDE simulation results of the vertical excitation energies of formaldehyde with different time lengths of the single Trotter step $t/M$. AEM stands for the excitation energies obtained by applying the algorithmic error mitigation, and Dotted horizontal lines indicate the CAS-CI excitation energies.} 
\end{figure}

\textbf{Vertical excitation energies of formaldehyde.} The two examples above deal with the energy gap between the lowest singlet and the lowest triplet states. Here, we focus on the vertical excitation energies of the spin-singlet excited states of formaldehyde (HCHO). We have calculated the excitation energies of three low-lying states, the $\mathrm{1\ ^1A_2}$, the $\mathrm{1\ ^1B_1}$, and the $\mathrm{2\ ^1A_1}$ states. As discussed in the Computational conditions section, these excited states can be approximated by the spin symmetry adapted ($\mathrm{2b_2 \rightarrow 2b_1}$), ($\mathrm{5a_1 \rightarrow 2b_1}$), and ($\mathrm{1b_1 \rightarrow 2b_1}$) one-electron excitations, respectively, from the $|\Phi_\mathrm{HF}\rangle$ state. Note that direct calculations of the excitation energies of these states have been reported by us using the BPDE algorithm.\cite{BPDE1st} The results are summarized in Figure~\ref{fig:fig7}. Again, the Trotter--Suzuki decomposition causes systematic error in the excitation energies, but the error can be efficiently removed by applying the AEM. The calculated excitation energies and those obtained from the BPDE simulations\cite{BPDE1st} are listed in Table~\ref{tab:table1}. Importantly, in the all excited states under study, the excitation energies obtained from the $N$-qubit QPDE in conjunction with the AEM agreed with the CAS-CI values with less than 0.006 eV = 0.138 kcal mol$^{-1}$ of deviation, which is smaller than that of the $\Delta E_{\mathrm{BPDE}}$ values listed in Table~\ref{tab:table1}. Note that the BPDE simulations\cite{BPDE1st} were carried out using $t/M = 0.50$. The excitation energies calculated by using the $N$-qubit QPDE and the BPDE methods with the same $t/M = 0.50$ value are very close to each other. The differences in the excitation energies between two calculations are mainly due to the shot noise in the BPDE simulations and the non-projective nature of the BPDE algorithm.

\begin{table}[t]
\caption{\label{tab:table1} Vertical excitation energies of formaldehyde calculated by using the $N$-qubit QPDE and the BPDE algorithms, and at the CAS-CI level of theory.}
\begin{tabular}{ccccccc}
\hline
\multirow{2}{*}{State} & \multicolumn{4}{c}{$\Delta E\mathrm{_{QPDE}/eV}$} & \multirow{2}{*}{$\Delta E\mathrm{_{BPDE}/eV}^{[b]}$} & \multirow{2}{*}{$\Delta E\mathrm{_{CAS-CI}/eV}$} \\
& $t/M = 1.25$ & $t/M = 1.00$ & $t/M = 0.50$ & AEM$^{[a]}$ \\
\hline
$\mathrm{1\ ^1A_2}$ & 4.942 & 5.092 & 5.293 & 5.360 & 5.297 & 5.359 \\
$\mathrm{1\ ^1B_1}$ & 10.135 & 10.268 & 10.469 & 10.530 & 10.466 & 10.525 \\
$\mathrm{2\ ^1A_1}$ & 11.053 & 11.270 & 11.588 & 11.686 & 11.603 & 11.692 \\
\hline
\end{tabular}
$^{[a]}$The energy gap obtained by applying the algorithmic error mitigation. $^{[b]}$Ref\cite{BPDE1st}. \\
\end{table}

\section{Conclusions}
In this work, we have developed a quantum phase difference estimation algorithm with $N_a$ of ancillary qubits for the direct calculations of energy gaps on a quantum computer. Although the number of ancillary qubits required to run the algorithm is larger than the previously proposed Bayesian inference-based implementations,\cite{BPDE1st, BPDE2nd, BPDEgrad, BPDE-FSS} the proposed approach is based on the projective measurement and therefore it is able to calculate the difference of energy eigenvalues of two electronic states with the approximated wave functions as inputs. The quality of the approximated wave functions only affects the success probability of the calculation, and the calculated energy gap value is independent of the input wave functions. The calculated energy gaps show systematic errors mainly caused by the Trotter--Suzuki decomposition. We demonstrated that the error is effectively reduced by applying the algorithmic error mitigation technique. For all the molecules being studied, the error-mitigated energy gap agreed to the full-CI or CAS-CI reference value with less than 1 kcal mol$^{-1}$ of deviations. Since sophisticated wave function preparation methods such as an adiabatic state preparation require deep quantum circuit, the single-shot protocol developed in this work is suitable to connect with such approaches. The $N$-qubit QPDE method with the input wave functions generated by using ASP is in progress and will be discussed in the forthcoming paper.

\begin{acknowledgement}
This work was supported by the Center of Innovations for Sustainable Quantum AI (Grant No. JPMJPF2221) from JST, Japan, Quantum Leap Flagship Program (JPMXS0120319794) from the MEXT, Japan, and Grants-in-Aid for Scientific Research C (21K03407) and for Transformative Research Area B (23H03819) from JSPS, Japan. Partial support by the PRESTO project "Quantum Software" (JPMJPR1914) from JST, Japan, is also acknowledged. 
\end{acknowledgement}


\bibliography{QPDE_refs}

\end{document}